\documentstyle[12pt]{article}
\begin{document}

\title{Lepton Spectra from $e^+e^-\to W^+W^-$\\in the BESS model}
\author{P. Poulose$^{1,}$\footnote{poulos@physik.rwth-aachen.de} , 
S. D. Rindani$^{2,}$\footnote{saurabh@prl.ernet.in}  and
L. M. Sehgal$^{3,}$\footnote{sehgal@physik.rwth-aachen.de}\\[3mm]
$^1${\em Institute of Theoretical Physics E, RWTH Aachen,}\\ 
{\em D-52056 Aachen,Germany}\\[2mm]
$^2${\em Theory Group, Physical Research Laboratory,}\\
{\em Navrangpura, Ahmedabad-380 009, India}}

\maketitle
\begin{abstract}
We investigate the reaction $e^+e^-\to W^+W^-,\;\;W^-\to l^-\bar\nu$
in a strong coupling scenario as implemented in the BESS model.  
Energy and angle spectra of the secondary lepton are calculated and
compared with the predictions of the Standard Model.  These spectra provide
a determination of
the fraction $f_0$ of longitudinally polarized $W$'s, and the backward
fraction of secondary leptons, $f_{back}$.  Assuming BESS
parameters allowed by present data, we give numerical estimates of the
effects to be expected at an $e^+e^-$ collider of energy $\sqrt{s}=500-
800$ GeV.
\end{abstract}

\def \to{\rightarrow}

\section{Introduction}

Since its introduction about three decades ago the Standard Model (SM) has
undergone stringent experimental tests, which it has so far been able to 
face successfully.
At the same time an important area of the theory is yet be tested.  
This is the mechanism that generates particle
masses in the SM, namely the Higgs mechanism.  
The idea of a scalar field which acquires a non-zero vacuum
expectation value is the central concept of the Higgs mechanism.  
This has the consequence of a physical scalar particle, the
mass of which the theory is unable to predict. Indirect mass bounds
obtained from high precision measurements at LEP and SLC are drawn with 
the assumption that this particle is elementary and sufficiently light.

An alternative scenario that has been discussed in the literature 
is the so-called 
``heavy-Higgs" limit \cite{veltman}.  The SM relation $m_H^2=\lambda\;v^2$
between the Higgs mass, $m_H$ and the electroweak scale, $v$, holds 
so along as $\lambda$, the quartic coupling of the scalar fields 
is perturbatively small.  In the
limit of large $\lambda$, the Goldstone bosons of the scalar sector become
strongly interacting, manifesting themselves as strongly interacting gauge
bosons.  
The elementary Higgs particle disappears from the spectrum, and instead 
various composite states appear as resonances in the $WW$ channel.
A specific approach in this direction goes under the name of
BESS (Breaking Electroweak Symmetry Strongly).  Reference \cite{bess} gives
a list of papers in this context.  In essence, BESS
borrows the idea of hidden local symmetry \cite{bala} of the non-linear 
$\sigma$-model, which has been applied with success to the understanding 
of pion-pion interactions at low energies \cite{bando}.  
By analogy with the pion-pion system, it is conjectured that there is  
a $\rho$-like resonance (gauge boson of the hidden 
local symmetry) in the electroweak interactions, which couples strongly 
to $WW$ pairs.  The mass of the new gauge boson, $m_V$, the gauge coupling 
of the new gauge sector, $g''$, and its direct fermionic coupling, 
$b$, are the parameters of 
the BESS model in addition to the parameters of the SM. A possible 
guideline for the parameter $g''$ is the analogous coupling $2\;g_{\rho\pi\pi}
\sim 2\;\sqrt{12\;\pi}\sim 12$ in the $\rho$-$\pi$-$\pi$ system.
A simple scaling of the $\rho$ resonance mass, $m_\rho$ to the electroweak
scale sets the typical mass of the new resonance. This gives
\[m_V=\frac{f_\pi}{v}\;m_\rho\sim 2\;{\rm TeV}\]
where $f_\pi$ is the pion decay constant.
Due to mixing of the standard gauge sector with the hidden gauge sector,
the new gauge bosons have induced coupling to the fermions 
even in the absence of the
direct fermionic coupling, $b$. Mixing also influences the couplings of
the standard gauge particles, the $Z$ and the $W$'s, to the fermions and
among themselves.  
A detailed description of the model can be found in the first two 
references in \cite{bess}.  

Precision measurements done at the LEP and SLC  colliders restrict the
parameter space of the BESS model.  
From the measured value of the radiative correction
parameter $\epsilon_3$ \cite{epsilon3}, Casalbuoni, 
{\em et al}, \cite{epsilon3a} obtained constraints in the ($g/g'',b)$ space.
Here $g$ is the standard weak coupling.
BESS contribution to $\epsilon_3$ is given in terms of the parameters as
\[\epsilon_3^{\sc bess}=-b+\left(\frac{g}{g''}\right)^2.\]
Considering this along with the SM corrections (obtained with the Higgs mass 
treated as a cut off) gives an allowed 
region of parameter space, defined by \cite{epsilon3a}
\[-\left(4.6^{-1.0}_{+0.5}\right)\times 10^{-3}\le
\epsilon_3^{\sc bess}\le \left(-0.5^{+1.0}_{-0.5}\right)\times 10^{-3}. \]
One can also consider the implications of the recent LEP2 data on the cross 
section of $WW$ pair production in the energy range of 183-207 GeV.  These
results agree with the SM prediction to 
an accuracy of about 2\% \cite{LEP2csec}.  Our
analysis shows that at c.m. energies around 200 GeV, the sensitivity
of the cross section to $g/g''$ is negligible as long as its value is 
less than about 0.1.  This enables us to put an upper limit of 0.01 on the
value of $b$.  Combining this with the constraint from $\epsilon_3$
restricts the value of $g/g''$ to be less than 0.05.  
We will accordingly consider parameter values in the general domain
$0\le b\le 0.01$ and $0\le (g/g'')\le 0.05$. 

The question of phenomenological interest is how these new gauge
bosons influence experimental observables. For example, the neutral 
gauge boson of the new gauge sector ($V^0$) behaves very much like 
the familiar $Z$ boson, interacting with different neutral currents.  
This can influence processes like $e^+e^-\to f\bar{f}$ or 
$e^+e^-\to W^+W^-$.  
Due to the comparatively strong coupling of the new gauge boson with $W$, 
pair production of $W$ in electron-positron collisions is a suitable 
candidate for testing the BESS model. Leptonic linear colliders at high
energies starting from 500 GeV are expected to be operational in the
foreseeable future.  
Previous studies \cite{casal:eeww} emphasise that the sensitivity 
to new physics is enhanced if one looks at the polarization of $W$.  
One way to do this is to look at the lepton spectra in 
$e^+e^-\to W^+W^-$ with $W^-\to l^-\bar\nu$.  
We will calculate the correlation of the lepton energy and the
lepton angle (in the lab frame), following the analysis of Koval'chuk {\em
et al.} \cite{koval}.  The energy spectrum turns out to be a function of
the diagonal elements of the $W^-$ spin density matrix ($f_0,\;\;f_+$ and
$f_-$), while the angular distribution contains additional information
involving also the non-diagonal elements.

In the next section we give expressions for the cross section and the
details of the observables considered.  In Sec. \ref{sec:results} 
we discuss the results, and make some concluding remarks in Sec. 
\ref{sec:conclu}.

\section{Signals from $e^+e^-\to W^+W^-$}

\begin{figure}[ht]
\vskip 5cm
\includegraphics{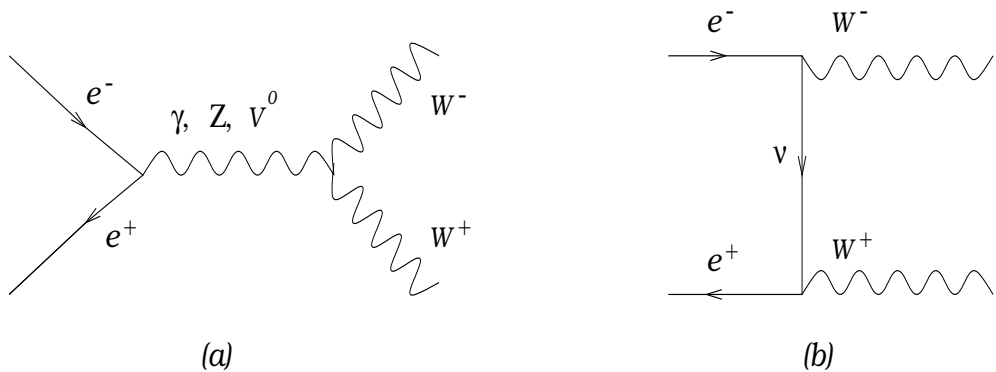}
\caption{Feynman diagrams contributing to the process 
$e^+e^-\to W^+W^-$.
}
\label{fig:feyndiag}
\end{figure}

To study the strongly interacting $W$'s in the context of the BESS model we
consider the process $e^+e^-\to W^+W^-$.  In addition to the SM channels,
this process gets contribution from an $s$-channel exchange of the
new gauge boson, $V^0$ (Fig.\ref{fig:feyndiag}).  

The relevant fermionic and gauge couplings are given in the Appendix.
The energy-angle correlation of the secondary lepton is calculated
following the procedure of Ref. \cite{koval}, using a Breit-Wigner form for
the $W^-$ propagator:

\begin{eqnarray}
\frac{d\sigma}{dx\;\;d\cos\theta_l}&=& \frac{3}{2}\;\;\frac{\alpha^2}{s} 
\;\;{\rm BR}(W^-\to e^-\bar\nu)\;\;A(s,x,\theta_l)\;\times \nonumber\\
&&\left[\tan^{-1}\left(\frac{m_W}{\Gamma_W}\right)+
\tan^{-1}\left(\frac{sx}{m_W\Gamma_W}-
\frac{s\tau}{m_W\Gamma_W(1-x)}\right)\right],
\label{eqn:ddistn}
\end{eqnarray}
where
\begin{eqnarray}
A(s,x,\theta_l)=C_sA_s+C_s'A_s'+C_{int}A_{int}+C_tA_t,
\nonumber
\end{eqnarray}
with
\begin{eqnarray}
A_s&=&-\frac{3}{2}-\tau-\frac{\tau}{x}+\frac{\tau^2}{x^2}+
\frac{x}{\tau}(1-x)\left(1+\frac{1}{4\tau}\right)+ \nonumber\\
&&\left(-\frac{5}{2}-\tau+3\frac{\tau}{x}-3\frac{\tau^2}{x^2}+
\frac{1}{2\tau}+\frac{x}{\tau}(1-x)\left(1-\frac{1}{4\tau}\right)\right)
\cos^2\theta_l \nonumber \\
A_s'&=&2\left(1+\frac{1}{4\tau}-2x-2\frac{\tau}{x}\right)\cos\theta_l 
\nonumber \\
A_{int}&=&-2\tau+\frac{x}{\tau}-2+
\frac{x}{2\tau}(1-x)\left(1+\frac{1}{2\tau}\right) + \nonumber\\
&&\left(1+\frac{1}{2\tau}-\frac{2\tau}{x}-2x\right)\cos\theta_l-
\left(1-\frac{1}{2\tau}\right)\left(1-\frac{x(1-x)}{2\tau}\right)
\cos^2\theta_l-\nonumber\\
&&R\;x^2\;\left(2+(cos\theta_l-\beta \cos\theta)\left(2-
\left(1+\frac{1}{\tau}\right)\beta\cos\theta+\cos\theta_l\right)\right)
\nonumber \\
A_t&=&\left(-2+\frac{2x}{\tau}+\frac{x(1-x)}{4\tau^2}\right)+
\frac{\cos\theta_l}{2\tau}+\left(1-(1-x)\frac{x}{2\tau}\right)
\frac{\cos^2\theta_l}{2\tau}- \nonumber \\
&&\frac{2}{\tau}x^2R(\beta \cos\theta-\cos\theta_l)\beta\cos\theta+
2 x^2aR^3(\beta\cos\theta-\cos\theta_l)^2.
\nonumber
\end{eqnarray}

Here $x=\frac{2E_l}{\sqrt{s}}$, where $E_l$ is the energy of the secondary
lepton in the $e^+e^-$ c.m. frame, $\sqrt{s}$ being the collider energy;
$\tau=\frac{m_W^2}{s}$;
$a=2\tau-1+\beta\cos\theta\cos\theta_l$, where $\cos\theta=\frac{1}{\beta}
(1-\frac{2\tau}{x})$ is the scattering angle of $W^-$, 
$\beta=\sqrt{(1-\frac{4m_W^2}{s})}$ is the velocity of $W^-$, 
and $\cos\theta_l$ is the polar angle of the secondary 
lepton, all in the c.m. frame; and $R^{-2}=4\tau^2+
(\beta\cos\theta-\cos\theta_l)(\beta\cos\theta-\beta^2\cos\theta_l)$.
BR$(W^-\to e^-\bar\nu)$ is the leptonic branching ratio of $W^-$.
The coefficients, $C_s,\;C'_s,\;C_t$ and $C_{int}$ involve various couplings, 
and are given in the Appendix. 

This formula gives the correlation of secondary lepton angle and energy.
The energy spectrum, integrated over all $\cos\theta_l$ is equivalent to
the angular distribution of the lepton in the rest frame of the $W^-$:

\begin{equation}
\frac{1}{\sigma} \frac{d\sigma}{d\cos\theta^*}=
\frac{3}{4}f_0\sin^2\theta^*+\frac{3}{8}f_+(1-\cos\theta^*)^2+
\frac{3}{8}f_-(1+\cos\theta^*)^2,
\label{eqn:fs}
\end{equation}
by virtue of the kinematical relation \cite{dicus}
\[E_l=\frac{\sqrt{s}}{4}\;(1-\beta\;\cos\theta^*).\]

\noindent
Here $\theta^*$ is the polar angle of the lepton in the rest frame of the
$W$ with $z$ axis along the boost direction.  $f_0$ gives
the fractional cross section of the longitudinal $W^-$, while $f_\pm$ give
that of the positive and negative helicity $W^-$'s.  
LEP2 has been able to measure the longitudinal fraction, 
$f_0$ with an accuracy of 5\% \cite{LEPf0}.  Although this does not 
restrict the parameters of the BESS model better than the total cross
section measurements at LEP2, the measurement of $f_0$ is of prime
importance in the search for new physics at high energies.

The energy distribution is studied in the context of BESS by 
Werthenbach and Sehgal in 
\cite{sehgal}.  One advantage of the energy-angle correlation given in 
Eq. \ref{eqn:ddistn} is that it enables us to study the effect of 
angular cuts, due to geometrical acceptance.  
For example an angular cut of $10^\circ$, as expected in the case of TESLA, 
distorts the energy spectrum in a way that can be computed using Eq.
\ref{eqn:ddistn}.

From the angular distribution of the lepton one can obtain, in particular, 
the backward fraction.  Since 
the $t$-channel $\nu$-exchange contribution peaks in the
forward direction, we expect the BESS-SM difference to be more pronounced 
in the backward hemisphere.
At high energies, the leptons are emitted more or less collinearly
with the $W$.  Therefore the backward fraction of the leptons is expected
to be a suitable observable to distinguish BESS from SM.  

These correlations and distributions can be used to obtain limits on the
BESS parameters that can be probed at future colliders. 
The secondary spectra, when 
combined with the primary observables studied in Ref. \cite{casal:eeww}, 
could help pin down the parameter space better. 
Our aim in this paper is to consider typical parameter values 
allowed by present experiments, and
study deviations expected at a collider running at different c.m. energies. 

We summarise the results of our analysis in the next section.

\section{Results}
\label{sec:results}

In our numerical analysis we consider the couplings restricted to 
$0\le b \le 0.01$ and $0.01\le (g/g'') \le 0.05$.
Mass of the new resonance is expected to be in the TeV range,
as suggested by the $\rho$-resonance mass in hadron-physics.  
We find that the observables are not very sensitive to the location of 
the resonance except rather close to the resonance.  
We give results for a 1 TeV resonance (which has a width  $\Gamma_V\sim 12$
GeV) and also for $m_V=2$ TeV (in which case $\Gamma_V\sim 350$ GeV). 
Our results are presented for two possible c.m. energies, 
500 GeV and 800 GeV as envisaged, e.g., for TESLA.

\subsection{\large \it Total Cross section:}
\begin{figure}[h]
\vskip 7.5cm
\includegraphics{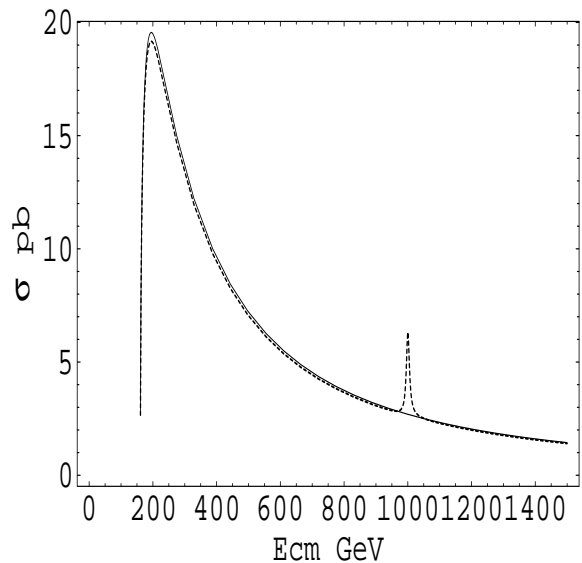}
\caption{Cross section of $e^+e^-\to W^+W^-$  against
        c.m. energy. Solid curve shows the SM value, while the
        dashed one represents the BESS value. BESS parameters
        considered are $b=0.01$, $g/g''=0.05$ and $m_V=1$ TeV.
        Intial beams are unpolarised.
}
\label{fig:csec}
\end{figure}

In Fig \ref{fig:csec} 
we recapitulate $\sigma(e^+e^-\to W^+W^-)$ in the SM and in the 
BESS scenario. Important feature is that the cross sections differ
significantly only in
the vicinity of the resonance. Numerical values for different choices of
($m_V,g/g'', b$) as well as electron polarization are given in
Table \ref{table:fback}.

\subsection{\large \it Energy Spectrum of the Secondary Lepton:}
The expression for energy distribution in terms of the polarization
fractions, as obtained from Eq. \ref{eqn:fs}, is

\begin{equation}
\frac{1}{\sigma} \frac{d\sigma}{d x}=\frac{2}{\beta^3}\;\left\{
\frac{3}{4}f_0\;\left(\beta^2-(1-2x)^2\right)+
\frac{3}{8}f_+\:\left(\beta-1+2x\right)^2+
\frac{3}{8}f_-\:\left(\beta+1-2x\right)^2\right\},
\label{eqn:en}
\end{equation}
where $x$ and $\beta$ have been defined earlier.

\begin{figure}[h]
\vskip 6.5cm
\includegraphics{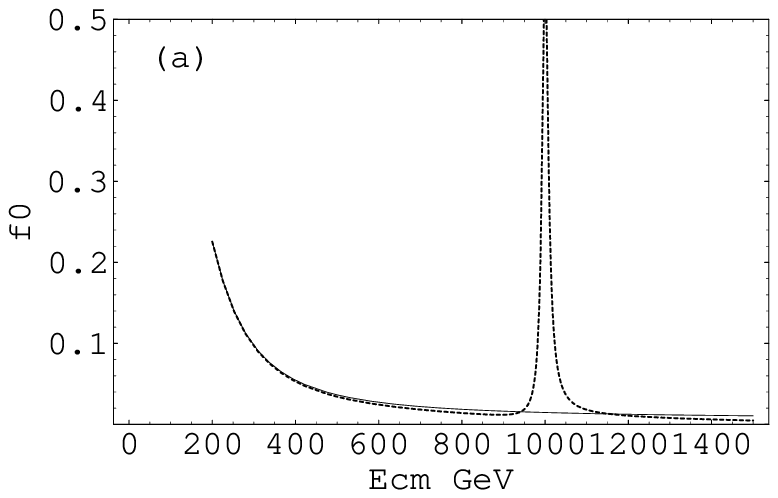}
\includegraphics{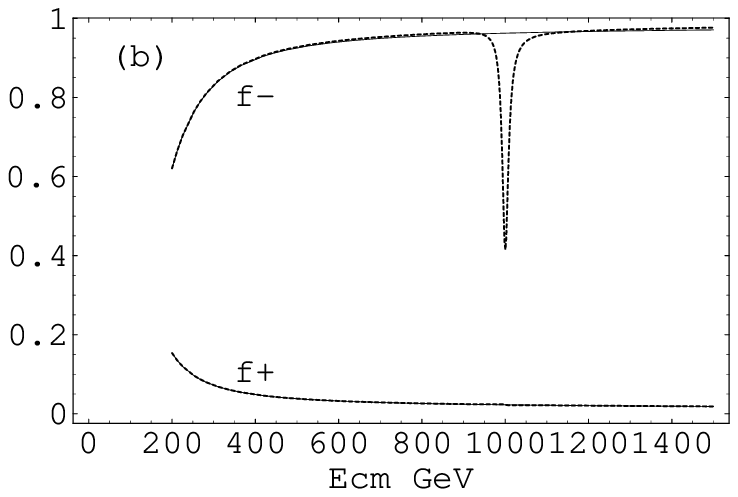}
\caption{Fractional cross sections of (a) longitudinal $W^-$
         and (b) transverse $W^-$ against c.m. energy. 
	Solid curve shows SM value and
        dashed curve represents BESS value with parameters
        $b=0.01$, $g/g''=0.05$ and $m_V=1$ TeV. Unpolarised
	initial beams are considered.
}
\label{fig:fs}
\end{figure}

Fig \ref{fig:fs} shows the polarization fraction of the $W$
against the c.m. energy.  The principal effects
occur in $f_0$ and $f_-$, especially when $\sqrt{s}$ is close to $m_V$.
The deviation at $\sqrt{s}=500$ GeV is about 6\% for parameter values 
of $b=0.01$ and $g/g''=0.05$. This is enhanced to a 25\% 
effect at 800 GeV.  Numerical values for the longitudinal $W$ fraction, 
$f_0$ are given in Table \ref{table:Prim} for $\sqrt{s}=500$ GeV and 
800 GeV.  Lowering the value of
$g/g''$ increases the sensitivity.  This is because of the fact that the
contributions proportional to $b$ and $g/g''$ compensate each other in the
range of parameters we are considering.  Thus for example, with $b=0.01$
and $g/g''=0.01$ deviation of $f_0$ is about 12\% at 500 GeV, 
which goes up to about 54\% at 800 GeV. 

We have also considered the case of polarized $e^-$ beams.
A priori, one would have thought that right-handed electron
beam would have helped, as there is no neutrino exchange contribution in
this case.  On the contrary, it turns out that a left-handed electron
beam is advantageous.  This is because the new vector boson couples to the
left-handed electrons much more strongly than to the right-handed ones.
For example, with a left-handed electron beam and 
an unpolarised positron beam, deviation of $f_0$ is larger than in
the case of unpolarised beams.  
At 800 GeV the deviation is improved from 25\% with unpolarized beams to
about 37\% with left-polarized electron beam for a parameter set, 
$b=0.01$ and $g/g''=0.05$. The corresponding improvement at 500 GeV is from
6\% to 8\%.

Table \ref{table:Prim} shows that differences between BESS and SM remain
detectable even if $m_V$ is raised from 1 to 2 TeV. In particular the
effect on $f_0$ is visible at 800 GeV for
some parameter values, even with a resonance at 2 TeV.

\begin{table}
\begin{center}
\begin{tabular}{c|ccll|cc|cc}
\hline
&&&&&\multicolumn{2}{c|}{}&\multicolumn{2}{c}{}\\
&&&&&&&&\\
&&&&&\multicolumn{2}{c|}{$\sqrt{s}=500$ GeV}&
\multicolumn{2}{c}{$\sqrt{s}=800$ GeV}\\[2mm]\cline{6-9}
&&&&&&&&\\
&$P_{e^-}$&$P_{e^+}$& $b$&$g/g''$&$\sigma^{\sc bess}/\sigma^{\sc sm}$
& $f_0^{\sc bess}/f_0^{\sc sm}$&$\sigma^{\sc bess}/\sigma^{\sc sm}$
& $f_0^{\sc bess}/f_0^{\sc sm}$\\[2mm]
\cline{1-9}
&&&&&&&&\\
$m_V=1$ TeV&0&0&0 &   0.05&    1.003&1.075&1.012&1.597\\
&&&0.01& 0.05&    0.977&0.937&0.975&0.754\\
&&&0&    0.01&    1.000&1.003&1.000&1.021\\
&&&0.01& 0.01&    0.975&0.875&0.969&0.463\\[2mm]\cline{2-9}
&&&&&&&&\\
&-1&0&0&    0.05&    1.003 &1.075&1.011&1.598\\
&&&0.01& 0.05&    0.977 &0.921&0.973&0.625\\
&&&0&    0.01&    1.000 &1.003&1.000&1.021\\
&&&0.01& 0.01&    0.975 &0.859&0.969&0.379\\[2mm]
\cline{1-9}\cline{1-9}
&&&&&&&&\\
$m_V=2$ TeV&0&0&0&0.05&1.002&1.061&1.005&1.253\\
&&&0.01& 0.05&0.977&0.940&0.976&0.786\\
&&&0&    0.01&1.000&1.002&1.000&1.009\\
&&&0.01& 0.01&0.975&0.888&0.973&0.620\\[2mm]\cline{2-9}
&&&&&&&&\\
&-1&0&0&    0.05&1.002&1.063&1.004&1.267\\
&&&0.01& 0.05&0.977&0.927&0.975&0.728\\
&&&0&    0.01&1.000&1.002&1.000&1.010\\
&&&0.01& 0.01&0.973&0.620&0.973&0.561\\[2mm]
\cline{1-9}
\end{tabular}
\end{center}
\caption{Ratios (BESS/SM) of total cross section and longitudinal fractions 
for different parameter values, without and with beam polarization.
($P_{e^-}=-1$ denotes left-handed electron polarization.)
}
\label{table:Prim}
\end{table}

\subsection{\large \it Angular Spectrum of the Secondary Lepton:}
\begin{figure}[h]
\vskip 8cm
\includegraphics{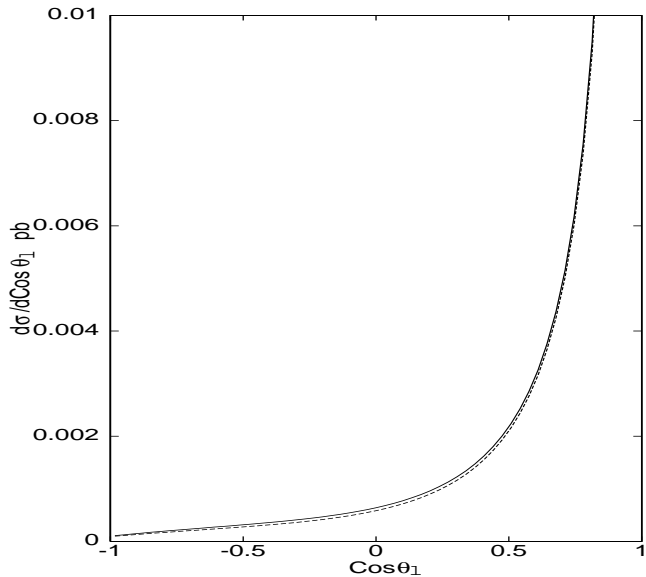}
\caption{Lepton angular distribution in the case of BESS (dotted line)
        and SM (solid curve). 
	BESS parameters are taken to be $b=0.01$,
        $g/g''=0.05$ and $m_V=1$ TeV. Beams are unpolarized and a c.m.
        energy of 800 GeV is considered. (Leptonic branching ratio not
	included.)
}
\label{fig:ctl}
\end{figure}

We obtain the $\cos\theta_l$ distribution by integrating out $x$ in 
Eq. \ref{eqn:ddistn}. The result is shown in Fig. \ref{fig:ctl}.  
The main effect is in the fraction of leptons 
at backward angles.  At 500 GeV, only 3 to 4\% of the decay leptons
are in the backward-hemisphere ($\cos\theta_l<0$).  This fraction changes
only by about 4 to 6\% in going from SM to BESS.
At 800 GeV the deviation becomes more significant ($\sim 21\%$).
These results are summarised in Table \ref{table:fback}.

\begin{table}
\begin{center}
\begin{tabular}{c|ccll|rc|rc}
\hline
&&&&&\multicolumn{2}{c|}{}&\multicolumn{2}{c}{}\\
&&&&&&&&\\
&&&&&\multicolumn{2}{c|}{$\sqrt{s}=500$ GeV}&
\multicolumn{2}{c}{$\sqrt{s}=800$ GeV}\\[2mm]\cline{6-9}
&&&&&&&&\\
&$P_{e^-}$&$P_{e^+}$&  $b$&$g/g''$&$\sigma$ pb
& $f_{back}$&$\sigma$ pb
& $f_{back}$\\[2mm]
\cline{1-9}
&&&&&&&&\\
$m_V=1$ TeV&0&0&\multicolumn{2}{c|}{{\rm S.M.}}&7.144&0.034&3.713&0.024\\[2mm]
&&&0 & 0.05&7.165&0.036&3.758&0.029 \\
&&&0.01& 0.05&6.981&0.033&3.620&0.022\\
&&&0& 0.01&7.144& 0.034& 3.715&0.024\\
&&&0.01& 0.01&6.964&0.032&3.600&0.019\\[2mm]\cline{2-9}
&&&&&&&&\\
&-1&0&\multicolumn{2}{c|}{{\rm S.M.}}&14.231&0.032&7.407&0.022\\[2mm]
&&&0& 0.05&14.269&0.033&7.486 &0.027\\
&&&0.01& 0.05&13.901&0.031&7.210&0.019\\
&&&0& 0.01&14.232&0.032 & 7.410&0.023\\
&&&0.01&    0.01&13.871&0.030&7.180&0.017\\[2mm]
\cline{1-9}
&&&&&&&&\\
$m_V=2$ TeV&0&0&\multicolumn{2}{c|}{{\rm S.M.}}&7.144&0.034&3.713&0.024\\[2mm]
&&&0 & 0.05&7.160&0.036&3.731&0.026\\
&&&0.01& 0.05&6.982&0.033& 3.622&0.022\\
&&&0& 0.01&7.144&0.034&3.714&0.024\\
&&&0.01& 0.01&6.968&0.032&3.611&0.020\\[2mm]\cline{2-9}
&&&&&&&&\\
&-1&0&\multicolumn{2}{c|}{{\rm S.M.}}&14.231&0.032&7.407&0.022\\[2mm]
&&&0& 0.05&14.261&0.033&7.441&0.025\\
&&&0.01&0.05&13.904&0.031&7.223&0.020\\
&&&0&0.01&14.232&0.032&7.409 &0.022\\
&&&0.01& 0.01&13.879&0.030&7.204&0.019\\[2mm]
\cline{1-9}
\end{tabular}
\caption{Total $WW$ cross section and backward fraction in the SM and 
BESS model for different parameter values. (Leptonic branching ratio not
included in $\sigma$.)}
\label{table:fback}
\end{center}
\end{table}

\subsection{\large \it Energy-Angle Correlation:}
\begin{figure}
\vskip 6.5cm
\includegraphics{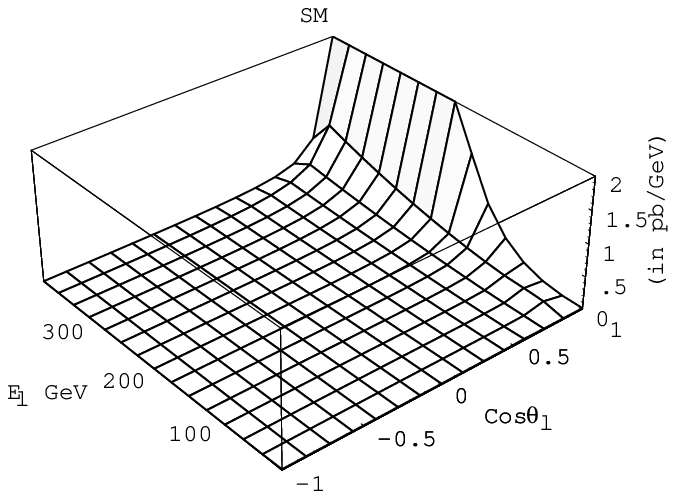}
\includegraphics{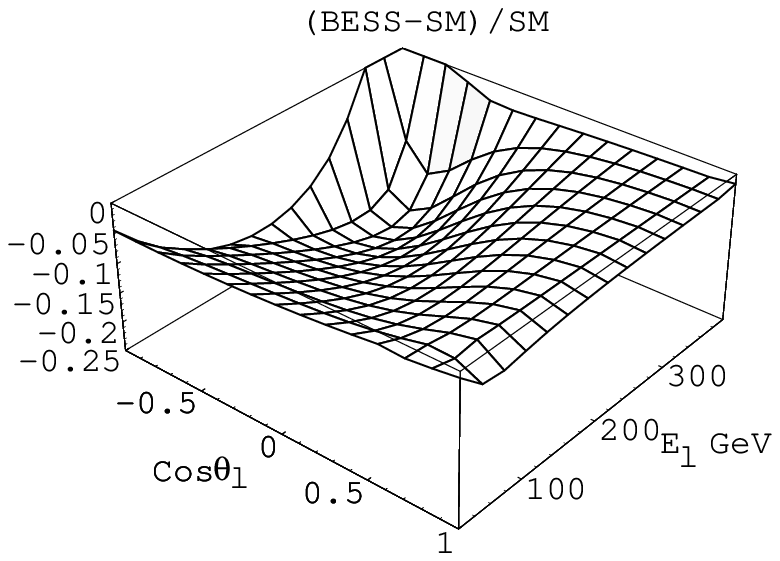}
\caption{(a) Correlation of leptonic energy and polar angle
        in the lab frame in the case of SM. 
	(b) Difference of the correlations in the BESS model
	and the SM. 
	BESS parameters used are $b=0.01$, $g/g''=0.05$ and 
	$m_V=1$ TeV.
        Intial beams are unpolarised and a c.m. energy of 800 GeV
        is considered in both the cases. (Leptonic branching ratio not
	included.)
}
\label{fig:ddsig}
\end{figure}

Correlation in the case of the SM,
$\frac{d\sigma^{\sc sm}}{dE_l\;d\cos\theta_l}$ 
is plotted in Fig.\ref{fig:ddsig} 
$(a)$, while Fig.\ref{fig:ddsig} (b) shows 
\( (\frac{d\sigma^{\sc bess}}{dE_l\;d\cos\theta_l}-
\frac{d\sigma^{\sc sm}}{dE_l\;d\cos\theta_l})/
(\frac{d\sigma^{\sc sm}}{dE_l\;d\cos\theta_l})\). 
At a c.m. energy of 800 GeV maximum deviation is about 25\% 
for a 1 TeV resonance with $b=0.01$ and $g/g''=0.05$. 
Corresponding value at 500 GeV is about 8\%.
However, the larger deviations tend to occur in kinematic regions where the
rate is small.

\begin{figure}
\vskip 6.5cm
\includegraphics{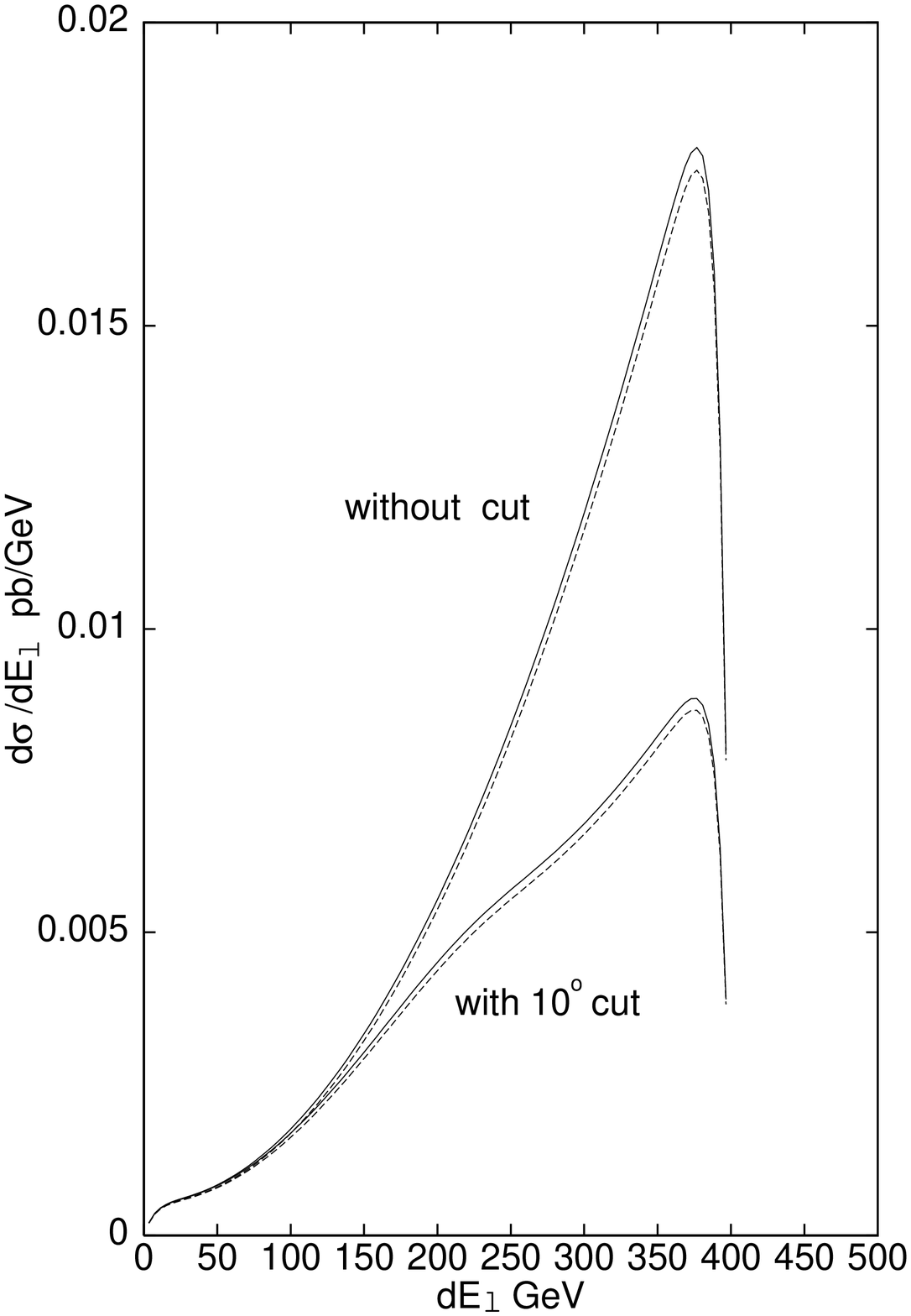}
\caption{Lepton energy distribution in the case of BESS (dotted lines) 
	and SM (solid curves). Upper set of curves is without any 
	angular cut, while the lower set is with a cut 
	$170^\circ\le \theta_l\le 10^\circ$  on the polar angle of the
	lepton. BESS parameters are taken to be $b=0.01$, 
	$g/g''=0.05$ and $m_V=1$ TeV. Beams are unpolarized and a c.m. 
	energy of 800 GeV is considered. (Leptonic branching ratio not
	included.)
}
\label{fig:en}
\end{figure}

One practical application of the expression in Eq. \ref{eqn:ddistn} is
that it allows us to calculate the effects of a geometrical cut that is
imposed by limited detector acceptance.  Such an angular cut results in a
distortion of the observed energy distribution, as shown in Fig.
\ref{fig:en}.

\subsection{\large \it High Energy Behaviour:}
\begin{figure}
\vskip 6.5cm
\includegraphics{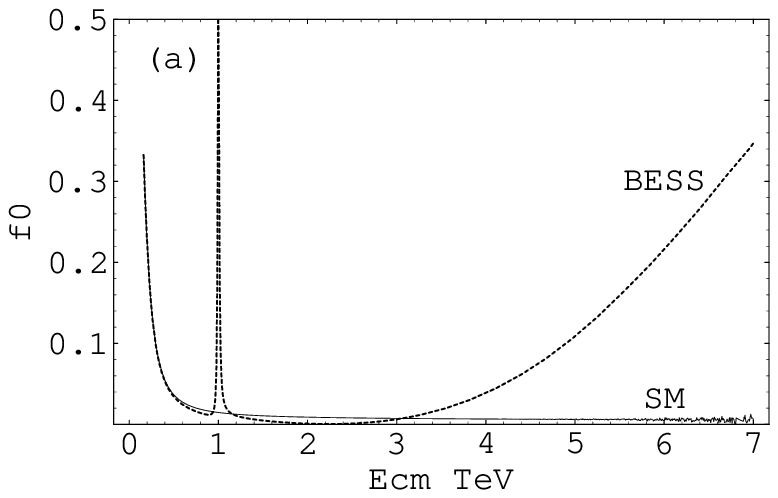}
\includegraphics{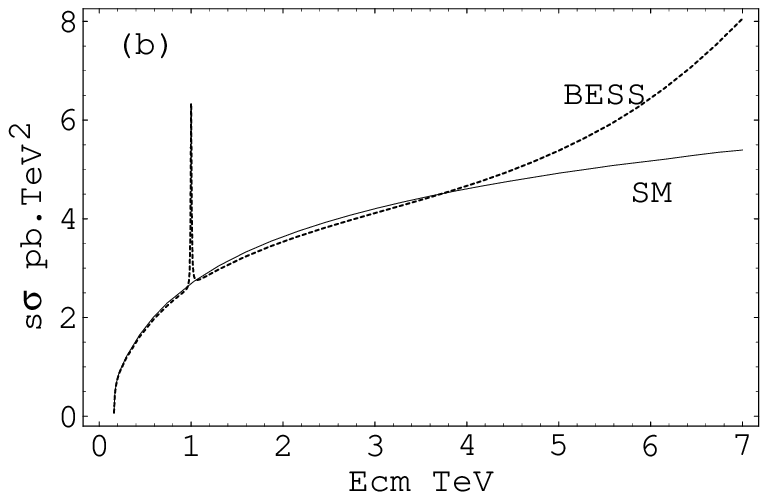}
\caption{(a) $f_0$ and (b) $s\sigma$ 
	against c.m. energy in the case of SM and BESS with
	parameters $b=0.01$, $g/g''=0.05$ and $m_V=1$ TeV. 
	Unpolarized beams are considered.
}
\label{fig:highen}
\end{figure}

One of the generic features of a strongly interacting Higgs sector is that
at sufficiently high energies, the longitudinal $W$ fraction dominates.  We
have checked this by looking at the behaviour of $f_0(\sqrt{s})$ at energies
far above the resonance.  As seen from Fig. \ref{fig:highen}$(a)$, the expected
dominance of $f_0$ sets in at multi-TeV energies.  Likewise, one expects
that in the BESS model (as contrasted with the SM) the cross section
$\sigma (e^+e^-\to W^+W^-)$ will deviate from the $\frac{1}{s}$ behaviour,
and ultimately become divergent, violating the unitarity limit, $\sigma <
12\pi/s$. This behaviour is also confirmed, as shown in 
Fig. \ref{fig:highen}$(b)$.
The features shown in Fig. \ref{fig:highen} are symptomatic of any 
non-standard model with a strongly interacting Higgs sector.

\section{Summary}
\label{sec:conclu}

We have studied the secondary lepton spectra coming from the $WW$ pairs
produced in $e^+e^-$ collisions to see the effect of the BESS model relative
to the SM. Our studies are complementary to earlier studies done on 
$e^+e^-\to W^+W^-$ which focussed on primary observables.  

With parameters allowed by low energy constraints from LEP and SLC we find
that a BESS type resonance in the 1 to 2 TeV region can produce small
effects in observables measured at $e^+e^-$ energies of $\sqrt{s}=500$ to
$800$ GeV.  These effects occur, in particular, in the longitudinal
helicity fraction $f_0$, which may be obtained from the energy spectrum of the
secondary lepton (see Eq. \ref{eqn:en}).  They also appear in the fraction
of secondary leptons produced in the backward hemisphere. A typical effect
is a change in the value of $f_0$ from 3\% in SM to 4\% in BESS.
Information on lepton spectra can eventually be incorporated in an analysis
such as that performed in Ref. \cite{casal:eeww} in order to delineate the
parameter space $(g/g'',b)$ of the BESS model.

\vskip 1cm
\noindent
{\large \bf Acknowledgements}\\[2mm]
One of us, P.P. wishes to thank the Humboldt Foundation for a
Post-doctoral Fellowship, and the
Institute of Theoretical Physics, RWTH Aachen for the
hospitality provided during this work.
L.M.S. would like to acknowledge a Visiting Scientist award from ICTP,
Trieste, enabling visits to the Physical Research Laboratory, 
Ahmedabad, where this project was initiated.

\vskip 1cm
\noindent
{\Large \bf Appendix}\\

\def \gww{g_{\gamma WW}}
\def \zww{g_{ZWW}}
\def \vww{g_{VWW}}
\def \enw{g_{e\nu W}}

\def \cvg{c^v_\gamma}
\def \cag{c^a_\gamma}
\def \cvz{c^v_Z}
\def \caz{c^a_Z}
\def \cvv{c^v_V}
\def \cav{c^a_V}

\def \cvgS{c^{v2}_\gamma}
\def \cagS{c^{a2}_\gamma}
\def \cvzS{c^{v2}_Z}
\def \cazS{c^{a2}_Z}
\def \cvvS{c^{v2}_V}
\def \cavS{c^{a2}_V}

The coefficients, $C$'s in Eqn \ref{eqn:ddistn} are given by

\begin{eqnarray}
C_s&=&\gww^2\;(\cvgS+\cagS)+s_Z^2\;\zww^2\;(\cvzS+\cazS)+
s_{V^2}\;\vww^2\;(\cvvS+\cavS)	\nonumber \\
&&+2\;s_Z\;\gww\;\zww\;(\cvg\cvz+\cag\caz)
+2\;s_V\;\gww\;\vww\;(\cvg\cvv+\cag\cav) \nonumber\\
&&+2\;s_Z\;s_V\;\zww\;\vww\;(\cvz\cvv+\caz\cav), \nonumber\\
C_s'&=&2\;(\gww^2\;\cvg\cag+s_Z^2\;\zww^2\;\cvz\caz
+s_{V^2}\;\vww^2\;\cvv\cav) \nonumber \\
&&+s_Z\;\gww\zww(\cvg\caz+\cag\cvz)+s_V\;\gww\vww(\cvg\cav+\cag\cvv)
\nonumber\\
&&+s_Z\;s_V\;\zww\vww(\cvz\cav+\caz\cvv)), \nonumber \\
C_{int}&=&\enw^2\;(\gww\;(\cvg-\cag)+s_Z\;\zww\;(\cvz-\caz)+
s_V\;\vww\;(\cvv-\cav)),\nonumber\\
C_t&=&\frac{\enw^4}{2},\nonumber
\end{eqnarray}
where the gauge couplings are given by
\begin{eqnarray}
\gww&=&1,\nonumber\\
\zww&=&\frac{\cos^2\phi}{\tan\theta_W}\;
\left(\frac{\cos\xi}{\cos\psi}+\tan\theta_W\;\tan\psi\;\sin\xi\right),
\nonumber\\
\vww&=&\frac{\cos^2\phi}{\tan\theta_W}\;
\left(\frac{\sin\xi}{\cos\psi}-\tan\theta_W\;\tan\psi\;\cos\xi\right)+
\frac{\cos\xi\;\sin^2\phi}{2\;\sin\theta_W}\;\frac{g''}{g}\nonumber
\end{eqnarray}
and the fermionic couplings by
\begin{eqnarray}
\enw &=&-\frac{1}{\sqrt{2x_W}\;(1+b)}
\left(\frac{\cos\phi}{\cos\psi}-\frac{b}{2}\frac{g}{g''}
\frac{\sin\phi}{\cos\psi}\right),\nonumber\\[1mm]
\cvg&=&1,\hskip5cm\cag=0, \nonumber\\
\cvz&=&-\frac{1}{4 \sin\theta_W\;\cos\theta_W }\;(-A+4\;B),
\hskip2cm\caz=\frac{A}{4 \sin\theta_W\;\cos\theta_W }, \nonumber\\
\cvv&=&-\frac{1}{4 \sin\theta_W\;\cos\theta_W }\;(-C+4\;D),
\hskip2cm\cav=\frac{C}{4 \sin\theta_W\;\cos\theta_W } \nonumber
\end{eqnarray}
with
\begin{eqnarray}
A&=&\frac{\cos\xi}{\cos\psi\;(1+b)}\;\left(1+b\;\sin^2\theta_W\;
\left(1-\frac{\tan\xi}{\tan\theta_W\sin\psi}\right)\right),\nonumber\\
B&=&\frac{\cos\xi}{\cos\psi}\;
\left(1-\frac{\tan\xi}{\tan\theta_W\sin\psi}\right)
\sin^2\theta_W,\nonumber\\
C&=&\frac{\sin\xi}{\cos\psi\;(1+b)}\;\left(1+b\;\sin^2\theta_W\;
\left(1+\frac{\cot\xi}{\tan\theta_W\sin\psi}\right)\right),\nonumber\\
D&=&\frac{\sin\xi}{\cos\psi}\;
\left(1+\frac{\cot\xi}{\tan\theta_W\sin\psi}\right)
\sin^2\theta_W.\nonumber
\end{eqnarray}

Here $\phi$, $\xi$ and $\psi$ are functions of BESS parameters as given below.
\[
\phi=-\frac{g}{g''},\;\;\;\;
\xi=-\frac{\cos 2\theta_W}{\cos\theta_W}\frac{g}{g''}\;\;\;\;\;{\rm and}
\;\;\;\;
\psi=2\;\sin\theta_W\;\frac{g}{g''},\]
where $g$ is the standard electroweak coupling.
The propagator factors are
\[s_Z=\frac{s}{s-m_Z^2},\;\;\;\; 
s_V=\frac{s\;(s-m_V^2)}{(s-m_V^2)^2+\Gamma_V^2m_V^2}\;\;\; 
{\rm and} \;\;\;\;
s_{V^2}=\frac{s^2}{(s-m_V^2)^2+\Gamma_V^2m_V^2}.\]

\end{document}